\documentstyle[12pt]{article}
\newcommand{\Db}{\overline D}
\newcommand{\chib}{\overline{\chi}}
\newcommand{\xib}{\bar{\xi}}
\newcommand{\z}{z_{12}}
\newcommand{\th}{\theta_{12}}
\newcommand{\thb}{\bar{\theta}_{12}}
\newcommand{\cT}{\cal T}
\newcommand{\cW}{\cal W}
\newcommand{\kp}{\kappa}
\newcommand{\gm}{\gamma}

\newcommand{\extraspace}{\addtolength{\abovedisplayskip}{2mm}
                        \addtolength{\belowdisplayskip}{2mm}
                        \addtolength{\abovedisplayshortskip}{2mm}
                        \addtolength{\belowdisplayshortskip}{2mm}}
\newcommand{\be}{\begin{equation}\extraspace}

\newcommand{\ee}{\end{equation}}

\newcommand{\bea}{\begin{eqnarray}\extraspace}
\newcommand{\beastar}{\begin{eqnarray*}\extraspace}
\newcommand{\eea}{\end{eqnarray}}
\newcommand{\eeastar}{\end{eqnarray*}}
\newcommand{\nonu}{\nonumber \\[2mm]}
\newcommand{\np}{Nucl.\ Phys.\ }

\newcommand{\cmp}{Comm.\ Math.\ Phys.\ }
\newcommand{\pl}{Phys.\ Lett.\ }
\begin{document}
\baselineskip 15pt

\begin{flushright} ITP-SB-93-18\\
April, 1993 \\
\end{flushright}
\vspace{4mm}

\begin{center}

{{\LARGE \sc Free Superfield Realization of \\
\vspace{2mm}
$ N=2$ Quantum Super $W_{3}$ Algebra}}\\

\vspace{2cm}

{{ \large \sc Changhyun Ahn}%
 \footnote{e-mail : ahn@max.physics.sunysb.edu}%
 \footnote{Researcher, Natural Science Institute, Yonsei University, Seoul,
  120-749, Korea}%
 \footnote{Address after Sept. 1, 1993 : Bogoliubov Theoretical Laboratory,
  JINR, Dubna, Head Post Office, P.O. Box 79, 101000, Moscow, Russia}} \\

\vspace{7mm}

{\it Institute for Theoretical Physics \\
     State University of New York at Stony Brook \\
     Stony Brook, NY 11794-3840} \\

\vspace{2cm}

\end{center}

\baselineskip=18pt

\indent

We rewrite $ N=2$ quantum super $W_{3}$ algebra, a nonlinear extended $N=2$
super Virasoro algebra, containing one additional primary superfield of
dimension $2$ which has no $U(1)$ charge, besides the super stress energy
tensor of dimension $1$ in $N=2$ superspace. The free superfield realization of
this algebra is obtained by two $N=2$ chiral fermionic superfields of dimension
$1/2$ satisfying $N=2$ complex $ U(1)$ Kac-Moody algebras.

\vfill

\newpage


\baselineskip=18pt

\setcounter{equation}{0}
\indent
Recently, Bershadsky, Lerche, Nemeschansky and Warner \cite{blnw} have
constructed that the $BRST$ currents for a non-critical $W_{3}$ string, minimal
$W$-matter model coupled to $W_{3}$-gravity, satisfy $ N=2$ super $W_{3}$
algebra \cite{ro,ny} and the general $BRST$ currents for non-critical
$W_{n}$-strings can be obtained by hamiltonian reduction \cite{bo} of the
affine super Lie algebra $SL(n|n-1)$.

Using the Polyakov's soldering procedure \cite{po} based on $ SL(3|2)$, the
'classical' $N=2$ super $ W_{3}$ algebra \cite{lprsw} in component formalism
has been studied explicitly. The other types of classical $N=2$ super $W_{3}$
algebra have been found in many literatures \cite{cn2w,ny,ek}. Ivanov and
Krivonos \cite{ik} have presented this 'classical' $N=2$ super $W_{3}$ algebra
\cite{lprsw} in
 $N=2$ superspace in terms of two supercurrents, found its Feigin-Fuchs type
representations by investigating two chiral $U(1)$ Kac-Moody supercurrents, and
obtained $N=2$ super Boussinesq equations.

Romans has determined the full structure of $N=2$ quantum super $W_{3}$
algebra\cite{ro} in component formalism by requiring Jacobi identities for
various (anti-) commutators.(See also \cite{od}). Other type of $N=2$ quantum
super $W_{3}$ algebra by one additional primary superfield of dimension $3/2$,
having non-vanishing
$U(1)$ charge, and its charge conjugate has been constructed \cite{swa}. And in
\cite{bl} all $N=2$ quantum super $W_{3}$ algebras up to dimension $5/2$ of the
additional generator have been discussed.
Very Recently \cite{ito}, the free field realization \cite{fz} of $ N=2$
quantum super $W_{3}$ algebra was  obtained from the $N=1$ super Miura
transformation \cite{ek} associated with  $SL(3|2)$, in component formalism and
found that the results are consistent with those \cite{ro}.

In this paper, we present $ N=2$ quantum super $W_{3}$ algebra \cite{ro} in $
N=2$ superspace and construct explicitly two supercurrents in terms of two
chiral superfields of dimension $1/2$ and finally the free superfield
realization of this algebra is obtained by chiral superfields of dimension $0$.

The $ N=2$ superconformal algebra \cite{ad,dpz,dsn2} is generated by the $ N=2$
super stress energy tensor ${\cT} (Z)$ of dimension $1$ written as
\bea
{\cT} (Z)=-J (z)+\frac{i}{\sqrt{2}} [\theta G^{-} (z)+\bar{\theta} G^{+}
(z)]-\theta \bar{\theta} T (z).
\eea
We follow the conventions of ref. \cite{dpz,swa}. $Z= (z,\theta,\bar{\theta})$
is a complex supercoordinate, $ J(z)$ is a $U(1)$ current of dimension $1$,
$G^{+}(z)$ and $G^{-}(z)$ are fermionic currents of dimension $3/2$ and $T(z)$
is the stress energy tensor of dimension $2$ \cite{ro}.

The $N=2$ superconformal algebra is represented by the operator product
expansion ( OPE ),
\bea
{\cT} (Z_{1}) {\cT} (Z_{2})=\frac{1}{\z^2} \frac{c}{3}-\left[ \frac{\th
\thb}{\z^2} +\frac{\th}{\z} D -\frac{\thb}{\z} \Db +\frac{\th \thb}{\z}
\partial \right] {\cT} (Z_{2})
\label{eq:TT}
\eea
where $c$ is the central charge and
\bea
\th = \theta_{1} - \theta_{2},\;\;\; \thb =
\bar{\theta}_{1}-\bar{\theta}_{2},\;\;\; \z =
z_{1}-z_{2}-\frac{1}{2}(\theta_{1} \bar{\theta}_{2}+\bar{\theta}_{1}
\theta_{2}).
\eea
We work with complex spinor covariant derivatives
\bea
D=\frac{\partial}{\partial \theta}+\frac{1}{2} \bar {\theta}
\partial,\;\;\;\;\; \Db =\frac{\partial}{\partial \bar{\theta}}+\frac{1}{2}
\theta \partial \nonu
\eea
satisfying the algebra
\bea
\{ D, \Db\}=\partial\;(=\partial_{z}),
\eea
all other anticommutators vanish.

Let us now consider the superalgebra that is obtained by adding a primary
supercurrent $ {\cW} (Z)$ of dimension $2$ to the superconformal algebra. We
write
\bea
{\cW} (Z) = V (z)-\frac{i}{\sqrt{2}} [\theta U^{-} (z)+\bar{\theta} U^{+}
(z)]+\theta \bar{\theta} W (z),
\eea
where $ V(z), U^{+}(z), U^{-}(z), \mbox{and}\; W(z)$ are primary fields of
dimension $2,\; 5/2,\; 5/2,$ and $3$ respectively.
The fact that ${\cW} (Z)$ is a primary superfield of dimension $2$ is expressed
by the OPE
\bea
{\cT} (Z_{1}) {\cW} (Z_{2})=-\left[ \frac{\th \thb}{\z^2} 2 +\frac{\th}{\z} D
-\frac{\thb}{\z} \Db +\frac{\th \thb}{\z} \partial \right]{\cW} (Z_{2}).
\label{eq:TW}
\eea

As a next step we consider the OPE $ {\cW} (Z_{1}) {\cW} (Z_{2})$ which can be
reexpressed in $N=2$ superspace completely. The following result corrects some
misprints in the expressions given in the ref. \cite{ro}.
\bea
& &{\cW} (Z_{1}) {\cW} (Z_{2})  =  \frac{1}{{\z}^4} \frac{c}{2}-\frac{\th
\thb}{{\z}^4} 3 {\cT} (Z_{2})-\frac{\th}{{\z}^3} 3 D {\cT} (Z_{2})\nonu
& & +\frac{\thb}{{\z}^3} 3 \Db {\cT} (Z_2)-\frac{\th \thb}{{\z}^3} 3 \partial
{\cT} (Z_{2})\nonu
& & +\frac{1}{{\z}^2} \left[ 2 {\kp} {\cW}-\frac{3}{(c-1)}
 {\cT}{ \cT}+\frac{c}{(c-1)}
[ D, \Db ] {\cT} \right] (Z_{2})\nonu
& & +\frac{\th}{{\z}^2} \left[ \kp D {\cW}-\frac{3}{(c-1)}
 {\cT} D {\cT}-
\frac{(2c-3)}{(c-1)} \partial D {\cT} \right] (Z_{2}) \nonu
& & +\frac{\thb}{{\z}^2} \left[ \kp \Db {\cW}-\frac{3}{(c-1)} {\cT} \Db
{\cT}+\frac{(2c-3)}{
(c-1)} \partial \Db {\cT} \right] (Z_{2})\nonu
& & +\frac{\th \thb}{{\z}^2} \left[
-\frac{3}{4} \gm ( 4c^3-11c^2+6c+36 )
\partial D \Db
{\cT}
\right. \nonu
& & -\frac{3}{2} \gm (8c^2-9c+36 ) {\cT} [ D, \Db ]
 {\cT}+9 \gm c( 5c-12 ) D
{\cT} \Db {\cT} \nonu
& & +9 \gm (4c+3) {\cT} {\cT} {\cT}-\frac{3 \kp (c-8) }{2(5c-12)} [ D, \Db ]
{\cW}
\nonu
& &\left.-\frac{42 \kp}{(5c-12)} {\cT} {\cW}-\frac{3}{4} \gm (
4c^3+19c^2-66c+36 ) \partial \Db D {\cT}
\right] (Z_{2}) \nonu
& & +\frac{\th}{\z} \left[
  \gm (-\frac{3}{
2} c^3-3c^2+45c-\frac{81}{2} ) \partial^2 D {\cT}
-3 \gm (c^2+24c-18) {\cT} \partial D {\cT}
\right. \nonu
& &+9 \gm (c^2+2c-3) D {\cT} \Db D {\cT} -9 \gm (2c^2-5c+3) D {\cT} D \Db {\cT}
\nonu
& &+9 \gm (4c+3) {\cT} {\cT} D {\cT} +\frac{3 \kp (c-1)(c-6) }{(c+3)(5c-12)}
\partial D {\cW}
\nonu
& &\left. -\frac{54 \kp (c-1) }{(c+3)(5c-12)} D {\cT} {\cW} +\frac{6 \kp (c-15)
}{(c+3)(
5c-12)} {\cT} D {\cW}
 \right] (Z_{2}) \nonu
& &+\frac{\thb}{\z} \left[
 \gm (\frac{3}{2} c^3+3c^2-45c+\frac{81}{2})
\partial^2
\Db {\cT}-3 \gm (c^2+24c-18) {\cT} \partial \Db {\cT} \right. \nonu
& & +9 \gm (c^2+2c-3) \Db {\cT} D \Db {\cT}-9 \gm (2c^2-5c+3) \Db {\cT} \Db D
{\cT}
\nonu
& & -9 \gm (4c+3) {\cT} {\cT} \Db {\cT}+\frac{3 \kp (c-1)(c-6) }{(c+3)(5c-12)}
\partial \Db
{\cW} \nonu
& &\left. +\frac{54 \kp (c-1) }{(c+3)(5c-12)} \Db {\cT} {\cW}-\frac{6 \kp
(c-15) }{(c+3)(
5c-12)} {\cT} \Db {\cW}
 \right] (Z_{2}) \nonu
& &+\frac{\th \thb}{\z} \left[
\frac{1}{2} \gm (-2c^3+16c^2-3c-18) \partial^3
{\cT}
\right. \nonu
& &+6 \gm c(5c-12) D {\cT} \partial \Db {\cT}-6 \gm c(5c-12) \Db {\cT} \partial
 D {\cT} \nonu
& &-3 \gm c(4c+3) D \Db {\cT} D \Db {\cT}+3 \gm c(4c+3) \Db D {\cT} \Db D {\cT}
\nonu
& &+18 \gm (4c+3) {\cT} {\cT} \partial {\cT}-\frac{\kp c(c-15)}{(c+3)(5c-12)}
\partial [ D,
\Db ] {\cW} \nonu
& &-\frac{18 \kp (c+6)}{(c+3)(5c-12)} {\cT} \partial {\cW}+\frac{6 \kp}{(c+3)}
D {\cT} \Db {\cW}+\frac{6 \kp}{(c+3)} \Db {\cT} D {\cW} \nonu
& &\left. -\frac{12 \kp (4c+3)}{(c+3)(5c-12)} \partial {\cT}
{\cW}-6 \gm
(c^2-3c+9) {\cT} \partial [ D, \Db ] {\cT}
 \right] (Z_{2}) \nonu
& &+\frac{1}{\z} \left[\kp \partial {\cW}-\frac{3}{(c-1)} {\cT}
\partial {\cT}+\frac{c}{2(c-1)} \partial [ D, \Db ] {\cT} \right] (Z_{2})
\label{eq:WW}
\eea
where\footnote{Multiple composite superfields are regularized from the left to
the
right.}\footnote{Normal ordered superfield product can be defined as
\cite{bs}.}
${\gm}$ and ${\kp}$ are given by \cite{ro}
\bea
\gm=\frac{1}{(c-1)(c+6)(2c-3)},\;\;\;
\kp=\frac{(c+3)(5c-12)}{\sqrt{2(c+6)(c-1)(2c-3)(15-c)}}
\eea

Thus we have rewritten $N=2$ quantum super $W_{3}$ algebra, eqs. (\ref{eq:TT}),
(\ref{eq:TW}) and (\ref{eq:WW}), for generic value of $c$ in $N=2$ superspace.
It is obvious that there exist a null superfield ${\Omega}(Z)$ for $c=12/5$
\cite{ro} which is contained in the discrete series of $N=2$ super Virasoro
algebra \cite{dsn2,dpz},
\bea
\Omega(Z) = {\cT} {\cW}(Z)-\frac{1}{5} [ D, \Db ] {\cW}(Z)
\eea
In fact, we can  easily see that the descendant superfields of $ \Omega  ( D
\Omega, \Db \Omega, \mbox{and} \;\partial \Omega )$ appear in the above OPE for
this $c=12/5$.
It would be interesting to investigate how the $N=1$ super $W_{3}$ algebra
\cite{ass} can be embedded in the $N=2$ quantum super $W_{3}$ algebra and
examine the algebra of twisted currents.

Our analysis at the quantum level is similar to the one presented at the
classical level in \cite{ik}. $N=2$ quantum super $W_{3}$ algebra can be
realized by two $N=2$ chiral fermionic superfields $\chi(Z), \xi (Z)$ of
dimension $1/2$ respectively,
\bea
D \chib = 0 = \Db \chi,\;\;\; D \xib = 0 = \Db \xi.
\eea
The defining OPE's are given by
\bea
{\chi} (Z_{1}) {\chib} (Z_{2}) =-\frac{1}{\z}+\frac{\th \thb}{\z^2}
\frac{1}{2},\;\;\; {\xi} (Z_{1}) {\xib} (Z_{2}) =-\frac{1}{\z}+\frac{\th
\thb}{\z^2} \frac{1}{2}
\eea

The $N=2$ super stress energy tensor can be obtained by having linear terms
\cite{yz,ik}
\bea
{\cT} (Z) =\left[ \chi \chib+\xi \xib+\sqrt{\frac{c-6}{6}}  \Db \chib
-\sqrt{\frac{c-6}{6}} D \chi \right] (Z)
\label{eq:T}
\eea
which satisfies the OPE (\ref{eq:TT}).
In order to extend the Feigin-Fuchs construction to the higher dimension
supercurrent ${\cW}(Z)$ we proceed as follows \cite{ik},
\bea
& & {\cW} (Z) =  \left[ c_{1} \partial D {\xi}+c_{2} \partial {\Db}
{\xib}+c_{3} \partial {\xi} {\chib}+c_{4} \partial {\xi} {\xib}+c_{5} \partial
{\xib} {\xi}+c_{6} \partial {\xib} {\chi}+c_{
7} \partial {\chi} {\chib} \right. \nonu
& &+c_{8} \partial {\chib} {\xi}+c_{9} {\xi} {\xib} {\chi} {\chib}+c_{10} D
{\xi} D {\chi}+c_{11} D {\xi} {\Db} {\chib}+c_{12} D {\xi} D {\xi}+c_{13} D
{\xi} {\Db} {\xib}  \nonu
& &+c_{14} {\Db} {\xib} D {\chi}+c_{15} {\Db} {\xib} {\Db} {\chib}+c_{16} {\Db}
{\xib} {\Db}
{\xib}+c_{17} D {\xi} {\xi} {\chib}+c_{18} D {\xi} {\xib} {\chi}+c_{19} D {\xi}
{\chi} {\chib} \nonu
& &+ c_{
20} D {\xi} {\xi} {\xib}+c_{21} D {\chi} {\xi} {\chib}+c_{22} D {\chi} {\xib}
{\chi}+c_{23} D {\chi} {\xi} {\xib}+c_{24} {\Db} {\xib} {\xi} {\chib}+c_{25}
{\Db} {\xib} {\xib} {\chi} \nonu
& &+c_{26} {\Db} {\xib} {\chi} {\chib}+c_{27} {\Db} {\xib} {\xi} {\xib} +c_{28}
{\Db} {\chib} {\xi} {\chib}+c_{29} {\Db} {\chib} {\xib} {\chi}+c_{30} {\Db}
{\chib} {\xi} {\xib}+d_{1} \partial {\chi} {\xib} \nonu
& & +d_{2} \partial {\chib} {\chi}+d_{3} \partial D
{\chi}  +d_{4} \partial {\Db} {\chib}+d_{5} D {\chi} {\chi} {\chib}+d_{6} \Db
{\chib} {\chi} {\chib}+d_{7} D {\chi} D {\chi} \nonu
& & \left. +d_{8} D {\chi} {\Db} {\chib}+d_{9} {\Db} {\chib} {\Db} {\chib}
\right] (Z)
\label{eq:W}
\eea
which has no $U(1)$ charge and the coefficients $ c_{1-30}$ and $d_{1-9}$
should be determined.

The above unknown $39$ coefficients reduce to the unknown $9$ coefficients
after imposing that ${\cW} (Z)$ should be a primary superfield (eq.
(\ref{eq:TW})). Finally all the coefficients can be expressed in terms of
$d_{1}$ or $c$ after we make the OPE calculations of ${\cW}(Z_{1})
{\cW}(Z_{2})$ (eq. (\ref{eq:WW})) which is very complicated, explicitly :
\bea
& &c_{1}=-\frac{\kp^2 (c-6)^{\frac{5}{2}}(c-1)^2}{\sqrt{6} (c+3)^2(5c-12)^2
d_{1}},\;\;\;
c_{2}=\sqrt{\frac{c-6}{6}} d_{1},\;\;\;c_{3}=-\sqrt{\frac{24}{c-6}} c_{1} \nonu
& &c_{4}=-\frac{\kp
(c^2+c+12)}{(c+3)(5c-12)},\;\;\;c_{5}=c_{4},\;\;\;c_{6}=-2d_{1} \nonu
& &c_{7}=-\frac{2 \kp (4c+3)}{(c+3)(5c-12)},\;\;\;c_{8}=\sqrt{\frac{6}{c-6}}
c_{1},\;\;\;c_{9}=-\frac{6 \kp}{(c+3)} \nonu
& &c_{10}=\sqrt{\frac{6}{c-6}} c_{1},\;\;\;c_{11}=\sqrt{\frac{6}{c-6}}
c_{1},\;\;\; c_{12}=\frac{\kp^3 (c-6)^3(c-1)^3}{(c+3)^3(5c-12)^3 d_{1}^{2}}
\nonu
& &c_{13}=-\frac{\kp
(3c^2-29c+12)}{(c+3)(5c-12)},\;\;\;c_{14}=d_{1},\;\;\;c_{15}=d_{1} \nonu
& &c_{16}=\frac{(c+3)(5c-12) d_{1}^2}{\kp
(c-6)(c-1)},\;c_{17}=-\sqrt{\frac{24}{c-6}}
c_{12},\;c_{18}=\frac{\kp \sqrt{96(c-6)}(c-1)}{(c+3)(5c-12)} \nonu
& &c_{19}=-\frac{6}{c-6} c_{1},\;\;\;c_{20}=\frac{6}{c-6} c_{1},\;\;\;
c_{21}=-\frac{6}{c-6} c_{1} \nonu
& &c_{22}=-\sqrt{\frac{24}{c-6}} d_{1},\;\;\;c_{23}=\frac{\kp \sqrt{6(c-6)}
(c-8)}{(c+3)(5c-12)},\;\;\;c_{24}=c_{18} \nonu
& &c_{25}=-\sqrt{\frac{24}{c-6}} c_{16},\;\;\;c_{26}=\sqrt{\frac{6}{c-6}}
d_{1},
\;\;\;c_{27}=-\sqrt{\frac{6}{c-6}} d_{1} \nonu
& &c_{28}=-\frac{12}{c-6} c_{1},\;\;\;c_{29}=-\sqrt{\frac{6}{c-6}}
d_{1},\;\;\;c_{30}=-c_{23} \nonu
& &d_{2}=c_{7},\;\;\;d_{3}=-\sqrt{\frac{c-6}{24}}
c_{7},\;\;\;d_{4}=-\sqrt{\frac{c-6}{24}} c_{7} \nonu
& &d_{5}=-\frac{7 \kp
\sqrt{6(c-6)}}{(c+3)(5c-12)},\;\;\;d_{6}=-d_{5},\;\;\;d_{7}=-\sqrt{\frac{c-6}{24}} d_{5} \nonu
& &d_{8}=\frac{\kp (c+48)}{(c+3)(5c-12)},\;\;\;d_{9}=-\sqrt{\frac{c-6}{24}}
d_{5}.
\eea
It is worth noting that the difference between the above results and those
\cite{ik} is the appearance
of
\bea
c_{7}, \;d_{2}, \;d_{3}, \;d_{4}, \;d_{5}, \;d_{6}, \;d_{7}, \;d_{8}\;
\mbox{and}\; d_{9}
\eea
terms which vanish in the limit $c \rightarrow \infty $.
To get the free superfield realization, we introduce the following
representation :
\bea
{\chi} = \Db {\bar{\Lambda}},\;\;\; {\chib}= D {\Lambda},\;\;\;
\frac{\xi}{d_{1}}= \Db {\bar {\Phi}}, \;\;\; {d_{1}} {\xib}=D {\Phi}
\label{eq:rep}
\eea
where ${\Lambda}, {\Phi}$ are the free chiral $ N=2$ superfields of dimension
$0$.
\bea
& &D {\bar {\Lambda} } = \Db {\Lambda}=0,\;\;\; D {\bar {\Phi}} = \Db {\Phi}=0
\nonu
& &{\Lambda} (Z_{1}) {\bar \Lambda} (Z_{2})=-\ln {\z}-\frac{\th \thb}{\z}
\frac{1}{2},\;\;\; {\Phi}(Z_{1}) {\bar \Phi}(Z_{2})=-\ln {\z}-\frac{\th
\thb}{\z} \frac{1}{2}
\eea
Therefore, we have the free field realization for supercurrents, eqs.
(\ref{eq:T}), (\ref{eq:W}) with eq. (\ref{eq:rep}).

The results obtained can be summarized as follows. We presented $ N=2$
manifestly supersymmetric quantum $W_{3}$ algebra, obtained two
supercurrents${\cT}(Z),{\cW}(Z) $ in terms of chiral superfields ${\chi} (Z),
{\xi} (Z)$
and this algebra was realized by free chiral superfields ${\Lambda} (Z), {\Phi}
(Z)$.

It would be interesting to
study, at the quantum level, for higher super extension of $W_{3}$ algebra
\cite{zamo}, for example, the classical $N=4$ super $W_{3}$ algebra \cite{pe}
constructed by the dual formalism. It is an open problem how to construct the
degenerate representation theory \cite{fz} (Vertex operators and Verma module )
from the above explicit form of free superfield realization and interprete the
relations between our results and those given in \cite{ito}.

\vspace{15mm}
I would like to thank M. Rocek for encouragement. This work was supported in
part by grant NSF PHY 9211367.

\end{document}